\documentclass[ps4pdf,showpacs,aps,prd,twocolumn,nofootinbib,superscriptaddress]{revtex4}
\usepackage{amsfonts,amsmath,amssymb,amsbsy,graphics,graphicx, epsfig}
\usepackage[colorlinks,hypertex]{hyperref}
\hypersetup{urlcolor=red,citecolor=blue}

\begin{document}
\title{Modulation Field Induces Universe Rotation}

\author{Chien Yu Chen}
\email{d9522817@phys.nthu.edu.tw}
\affiliation{\small Department of Physics, National Tsing-Hua University, Hsinchu 300, Taiwan}

\begin{abstract}
In this paper, we consider a time dependent module field on spacetime extension without modifying commutative relation on noncommutative quantum plane. The significant idea is that $Lorentz$ symmetry is conserved in module and unmodule coordinate. We focus on the redefinition of spacetime structure without considering noncommutative bosonic gas in deforming the product between fields. Which the null vector is a vector on orthogonal $D$ dimensional $Hilbert$ spacetime. In $Riemann$ geometry, the equation of motion is deformed from an induced rotation. Particle field survives on the state composed by two theoretical assumed $null$ vectors, one is commutative, another is anticommutative. In the point of view, neutrino and photon mass are produced by its shift, the rotated effect generates a horizon in redefining particle field.
\end{abstract}

\maketitle

\section{Introduction}
Noncommutative universe is a $Lorentz$ and parity violated field theory. The general idea is to modify the commutative relation by imposing a constant and uniformed background~\cite{Seiberg:1999vs} in Dirac-Born-Infeld string action and noncommutative quantized gauge~\cite{Martin:2002nr}. The paper in $Carroll$ et. al~\cite{Carroll:2001ws} mentioned that we can roughly consider an effective antisymmetric tensor on an special universe framework\footnote{Editor suggested in response report}. The possibility of spontaneous $Lorentz$ violation is manifestly involved into the couplings of standard theory. This paper we consider the graviton effects in rendered into noncommutative via a coupling to the converted background tensor field with eqn 10 containing in $vieblein$ representation of metric.

The decided features of coordinate noncommutable effect is forwarded in the patch of,
\begin{equation}\label{eq1}
[\hat{x}^{\mu},\hat{x}^{\nu}]_{\star} = i\theta^{\mu\nu},
\end{equation}
where
\begin{equation}\label{eq2}
\theta_{\mu\nu} = \frac{1}{\Lambda^{2}_{C}}\begin{pmatrix}0&E_{1}&E_{2}&E_{3}\\-E_{1}&0&-B_{3}&B_{2}\\-E_{2}&B_{3}&0&-B_{1}\\-E_{3}&-B_{2}&B_{1}&0\end{pmatrix},
\end{equation}
the scale $\Lambda_{C}$ = $10^{19}$ GeV. The commutation relation is consistent with Schr$\ddot{o}$dinger algebra~\cite{Banerjee:2005zt}. It requires that even $Lorentz$ is automatically violated, but angular momentum conservation is still in a rule. In Eq.($\hyperref[eq1]{1}$), the pseudotensor $\theta_{\mu\nu}$ is a $Lorentz$ scalar. Each component inside the tensor will not be changed by $Lorentz$ transformation, $\theta_{\mu\nu}$ = $\Lambda_{\mu}^{~\alpha}\theta_{\alpha\beta}\Lambda^{\beta}_{~\nu}$. In $Lorentz$ violated $\theta_{\mu\nu}$ deformed spacetime, it will generate a parity violated effects in collider physics~\cite{Chen:2008mh}. In this paper, we consider a model with module extension of $Pancar\check{e}$ translation at $Planck$ scale. The expanded $Pancar\check{e}$ boost parameter is a time and space dependent one. The degrees of freedom in assuming that U(1) gauge parameter is the same as coordinate boost parameter. It is a similar way to select a $Pancar\check{e}$ gauge freedom from twisting commutative relationship. We choose spacetime gauge is consistent with U(1) gauge parameter by taking a constraint on null vector.

Separating null vector to commutative and anti-commutative property is in order to describe a general phenomenon of photon mass and asymmetric effects of observed universe. We can choose a preferred frame to ignore the anisotropic universe structure generated in module translation. Hence, whole of the universe will be almost parity conserved. On the contrast, if we stand on the free falling frame, the curved phenomenon will be an important role to correct observed results. Moreover, if we assume that background strength tensor is a random field with magnetic and electric field filled up into each cell, in which, the preferred direction is decided by a selected null vector.

Considering full degrees of freedom extension in null vector, it is composed by three unit orthogonal vectors with unequal length. Each component in $\theta_{\mu\nu}$ is a $Lorentz$ scalar with constant component translated between each coordinates within a surroundings of background $\theta_{\mu\nu}$ deformed photon gas~\cite{Barosi:2008gx, Bemfica:2005pz}. The profile of present universe is assumed that the background is a constant and uniformed. $Lorentz$ scalar tensor $\theta_{\mu\nu}$ and elliptic $null$ vectors are a source of parity violation in high energy collider experiments. $Lorentz$ invariance is required in $\theta_{\mu\nu}$ by containing an inner structure depends on its location.

In which, the plagued mixing of UV/IR divergence is a two dimensional nonlocal effects~\cite{Filk:1996dm}. The planar diagram is the interaction residing in the local spacetime structure, as for nonplanar diagram the process includes a non-local fluctuations in its vacuum perturbation. The corresponded phenomenon indicates to correct particle horizon surface shell. The mixture between each limit includes the contribution between UV and IR branes. As those branes are mixed in a special geometry. The time component of momentum including into integral with non-planar fluctuations is similar to consider a deformed structure connecting the fluctuations between space and time. It will produce $causality$ problem with the result of time is a function of space. In order to avoid the mixture, the perturbative parameter, $\theta_{\mu\nu}$, is restricted from the basis of without condensing time and space in nonlocal patch.

Noncommutative slice patch constrains from unitarity condition in setting $\theta_{0i}$ = 0, but, it is not very completely in constraining $causality$ problem. Whether the component in $\theta_{0i}$ = 0 or not, microcausality condition does not be completely cancelled in $\theta_{ij}\neq$ 0~\cite{Greenberg:2005jq}. In redefining $Schr\ddot{o}edinger$ symmetry commutative properties, taking the reparameterized operator into account~\cite{Banerjee:2005zt}, the general angular momentum in noncommutative spacetime will reduce to classical one in choosing background preferred coordinate.

As to the case without $\theta_{0i}$ components, the two parameters are chosen to be orthogonal to the angular momentum. Hence, the anomaly angular momentum distribution will be naturally modified~\cite{Colatto:2005fy}, and the $ABJ$ anomaly produced from noncommutative bosonic gas, with $\theta_{ij}$ components, under the product of $[~,J_{\theta}]_{\star}$ emerged into noncommutative departure coupling to commutative field~\cite{Banerjee:2005zq}. Otherwise, we show that geometric fluctuations either generates $ABJ$ anomaly in fermion action without taking account of $Moyal~Weyl$ product between each particle fields.

Building up module field on spacetime four vectors and considering an alternative matrix product, the other degrees in $Nambu~Gostone$ boson production is considered. Thereof, the background fluctuations, vacuum generates a mass term by eating ground state degrees of freedom~\cite{WQFT}. The rotated vector, $\theta^{\mu\nu}n_{\nu}$, imposes an extra degrees by choosing a preferred frame within global universe. In O(1) case, modulation field is a massless $Nambu~Goldstone$ boson, its energy momentum distribution is satisfied by Klein-Gordon equation. Which universe structure is merged in $\rho$, and gauged in bosonic condition. The degrees of freedom in equation of motion of modulation, $\partial_{\mu}\rho\partial^{\mu}\rho$ = 0, is decided by universe structure.

Extending graviton field into $Weyl~Moyal$ product~\cite{Bertolami:2002eq, Bertolami:2005hz, Buric:2006di, Rivelles:2002ez}, we consider a geometry with composing a symmetric and an antisymmetric graviton, and its departure with module translation
\begin{equation}\label{eq3}
d\hat{x}^{\mu}\to dx^{\mu} + \theta^{\mu\lambda}n_{\lambda}\mathcal{K}\rho,
\end{equation}
where $\rho$ is a time dependent $Nambu~Goldstone$ boson and $\mathcal{K}$ is particle canonical momentum on $null$ frame. Ref.~\cite{Brandt:2006ua}, the derivative is a product of module or $Nambu~Goldstone$ field $f(x)$. The content depends on its location and situation. We consider a Lorentz violated transformation in a non-$null$ state. The translation between each frame is under $vielbein$ rotation, $\hat{e}^{m}_{~\mu}$, from magnetic frame to $Einstein$ frame. The produced modification adds a freedom degrees in redefinition of spacetime geometry.

The constructed module model is similar working on $Pancar\check{e}$ group. It does not contain inverse $vieblein$ field unless we take off the shifted spacetime degrees of freedom by hand. In order to generate the $vieblein$ translation, some degrees have to be absorbed into graviton gauge freedom. $Nambu~Goldstone$ boson is produced by $Pancar\check{e}$ translation~\cite{Bluhm:2004ep}. Either, noncommutative modulation translation Eq.($\hyperref[eq3]{3}$), 10 $Nambu~Goldstone$ boson is produced by spacetime extension. If we impose a tensional geometry on vacuum and absorb 10 degrees into graviton field, the interaction between $Nambu~Goldstone$ field with graviton is considered. Due to traceless gauge, background field cannot stuff energy on universe vacuum. Phenomenologically, particle field in ground state with a extended vector $\theta_{\mu\nu}n^{\nu}$ will generate a mass term from rotated background universe structure.

\section{Modulation Field}
Dividing geometric tensor into symmetric and antisymmetric parts, the exchange from each coordinate frame, $dx^{\mu}$, are the same as in symmetric property. The antisymmetric tensor cannot be contributed on a symmetric spacetime. In assuming that null vector contains symmetric $n^{\mu}_{S}$, antisymmetric $n^{\mu}_{A}$ vectors and the production of $n^{\mu}\theta_{\mu\nu}n^{\nu}$ = 0. Whether those survive in elliptic frame or not, translation relation, Eq.($\hyperref[eq3]{3}$), between free falling frame and null frame will be inversible due to the inherent commutation relation in null vector and the equal length condition $n_{\mu}d\hat{x}^{\mu}$ = $n_{\mu}dx^{\mu}$. Some degrees of freedom have been absorbed into graviton gauge freedom parameter. We consider the geometry by adding an antisymmetric graviton. The anomaly in ward identity is attributed to the production between itself with antisymmetric $null$ vector.

Conformal translation, $\tilde{g}_{\mu\nu}$(x) = $\Omega$(x)$g_{\mu\nu}$(x), considers a maximum symmetry in massless field. Spacetime translation, in $Lorentz$ group $\tilde{x}^{\mu}$ = $\Lambda^{\mu}_{~\nu}x^{\nu}$ and in $Pancar\check{e}$ group $x'^{\mu}$ = $\Lambda^{\mu}_{~\nu}x^{\nu}$ + $\Lambda^{\mu}_{~\nu}a^{\nu}$, $a^{\mu}$ is a constant four vector, are fully at first order $conformal$ translation. It means that spacetime still includes an extension degrees of freedom from choosing the parameter $a^{\mu}$. The killing vector is also under $Lorentz$ group in order to preserve isometric map. However, the others breaking $Lorentz$ symmetry are also discussed~\cite{Cohen:2006ky, Kostelecky:1988zi, Kostelecky:2003fs}. The field moving along element geodesic $dx^{\mu}$ under the anisotropic one, that allows others freedom to generate the slight correction, $\Delta x^{\mu}$ = $d\hat{x}^{\mu}$ - $dx^{\mu}$ = $\hat{\xi}^{\mu}(x)$\footnote{The $\hat{\xi}^{\mu}(x)$ is associated with the $Planck~scale$ mass and slight shift inflationary phenomena.}. The original element line is symmetric in $Lorentz$ transformation, but $\xi^{\mu}$ does not has to be. $\Delta x^{\mu}$ contributes on the outside isotropic one, and the contribution is dependent on the component whether time or space. The preferred direction induces an inherent parity violation in universe, but the time direction is ignored by $vieblein$ and inverse $vieblein$ transformation.

Spontaneous $Lorentz$ violation in string theory by taking tachyon field in witten string effective potential~\cite{Kostelecky:1988zi}. In noncommutative field theory, imposing a constant background field on universe vacuum and considering an uniformed direction in isotropic spacetime. Under $Seiberg~Witten$ map~\cite{Seiberg:1999vs,  Martin:2002nr} the finial consequence is to amend $Pancar\check{e}$ parameter, $\xi^{\mu}$, on spacetime coordinate commutative relation. In the aspects of effective theory extension~\cite{Myers:2003fd}, modifying particle energy distribution by putting an 5 dimensional operator into lagrangian. On the empty vacuum, if no $Nambu~Goldstone$ field absorbed into tensional geometry, we cannot find out an inverse transformation between free falling frame and module field situation.

In CPT conserved theory, dispersion relation cannot modify photon energy momentum distribution. On the other hand, photon is still survived in IR region. By considering its mass term, we impose all the degrees of freedom into null state commutation relation. The symmetric part devotes into the correction of photon mass dispersion and antisymmetric part is contributed on anomaly term and graviton radiation. Hence, we discuss the assumption of constant and uniformed background field, the product length with photon and elliptic null frame is decided by the profile of unit elliptic ball. On the elliptic null frame, the random vector couples to momentum vector induces a parity violated phenomena from alternation of null vector.

Each vector is vividly imposed into elliptic frame by introducing a module field and unequal length of $null$ vector from spacetime extension. According to coordinate translation, $\hat{x}'^{\mu}$ = $\frac{\partial\hat{x}'^{\mu}}{\partial x^{\alpha}}x^{\alpha}$ or $\hat{g}'^{\mu\nu}=\frac{\partial\hat{x}'^{\mu}}{\partial x^{\alpha}}\frac{\partial\hat{x}'^{\nu}}{\partial x^{\beta}}\eta^{\alpha\beta}$, between non-background field and background field. The translation between $Riemann$ geometry and $Mincowski$ geometry is in order to diagonalize the metric of arbitrary form. Splitting the $vielbein$ field into two parts $E^{a}_{~\mu}$ = $e^{a}_{~\lambda}\hat{e}^{b}_{~\mu}\delta^{\lambda}_{b}$~\cite{Hebecker:2003iw}, the similar form is as follows
\begin{equation}\label{eq4}
\begin{split}      
d^{2}\hat{s}' &= \eta_{ab}d\hat{x}'^{a}\otimes d\hat{x}'^{b} = E^{a}_{~\mu}\otimes E_{a\nu}dx^{\mu}dx^{\nu}\\
              &= d\hat{x}^{0}d\hat{x}^{0} - (\delta_{ij} - h_{ij} - \mathcal{B}_{ij})d\hat{x}^{i}\otimes d\hat{x}^{j},\\
\end{split}
\end{equation}
where $E^{a}_{~\mu}\otimes E_{a\nu} =  \frac{1}{2}(E^{a}_{1\mu}E_{2a\nu} + E^{a}_{2\mu}E_{1a\nu})$ = $(\eta_{\alpha\beta} + h_{\alpha\beta} + \mathcal{B}_{\alpha\beta})\delta^{\alpha}_{a}\delta^{\beta}_{b}\hat{e}^{a}_{~\mu}\otimes\hat{e}^{b}_{~\nu}$, and $\hat{e}^{a}_{~\mu}\otimes\hat{e}^{b}_{~\nu}$ = $\frac{1}{2}(\hat{e}^{a}_{1\mu}\hat{e}^{b}_{2\nu} + \hat{e}^{a}_{2\mu}\hat{e}^{b}_{1\nu})$. In convention, the index $1$ and $2$ are in order to discriminate the assumed elliptic geodesic line. The translation of $\hat{e}^{a}_{~\mu} = \frac{d\hat{x}^{a}}{dx^{\mu}}$ = $\delta^{a}_{~\mu} + \theta^{a}_{~\lambda}n^{\lambda}\mathcal{K}\partial_{\mu}\rho$, and $\eta_{ab}e^{a}_{~\mu}e^{b}_{~\nu}$ = $\eta_{\mu\nu} + h_{\mu\nu} + \mathcal{B}_{\mu\nu}$, where $h_{\mu\nu}$ is a symmetric graviton and $\mathcal{B}_{\mu\nu}$ is an antisymmetric field~\cite{Hebecker:2003iw}. In physical concept, we have to choose an inertial frame for physical system. In general, however, proper time produces a matrix geodesic from producing with two observer vectors $\mathcal{U}^{\mu}$.

Transforming to the inertia frame by $vieblein$ field, containing a 10 degrees, dominates graviton's degree of freedom in general relativistic theory. Thus, we expand the geodesic line in proper time perturbative extension~\cite{Mashhoon:2008vr},
\begin{equation}\label{eq5}
A'^{\mu}(t,\vec{x}) = A^{\mu}(\tau,\vec{x}) + \frac{\delta}{\delta\tau}A^{\mu}(\tau,\vec{x})\delta\tau,
\end{equation}
the $\frac{\delta}{\delta\tau}A^{\mu}(\tau,\vec{x})$ term is an depiction of tensional geodesic. After rotation, diagonalizing the mixing matrix, the eigenvalue of rotation matrix is devoted into a change from the displacement of spheric shell into elliptic profile. The elliptic profile can be found in nucleus system by considering that nucleus contains two shell. Ref.~\cite{Bastrukov:2008fh} mentions that an elliptic phenomenon is produced by translating a soft Goldstone boson between two lays. The equation of motion of graviton spinor field and observer, it is standing on the frame moving along with objects.

In the elliptic universe, electric and magnetic field are survived in non-orthogonal site. Hence, the expectation value between the product of magnetic field and electric field should be an observable quantity. The anomaly phenomenon is revealed from fermion eigenstate perturbatively by axial current and two vector currents triangle diagram.

Gravitational radiation is coming from the interaction with background. By taking harmonic gauge, $\eta^{\alpha\beta}\Gamma^{\mu}_{~\alpha\beta}$ = 0, and TT gauge into account, the radiations between each graviton are omitted in the linear order purterbative correlation. The interaction with background field, gravitational radiation is an important feature to explore graviton wave. We get graviton property from detecting background energy flux. Particle is scattered from graviton radiation in the same mode interaction under helicity conservation. However, the field works in graviton spin wave background, its behavior is similar to work on a rotated surroundings. From the plane ware expansion, graviton field can be read as incoming and out going field
\begin{equation}\label{eq6}
h_{\mu\nu}(x) = e_{\mu\nu}e^{ik^{\mu}x_{\mu}} + e^{\star}_{\mu\nu}e^{-ik^{\mu}x_{\mu}},
\end{equation}
where the $vierbien$ field is constrained into radiation condition, $k_{\mu}e^{\mu}_{~\nu}$ = $\frac{1}{2}k_{\nu}e^{\mu}_{~\mu}$, and the gauge constraint, $k^{\mu}k_{\mu}$ = 0. We get a form of modifying graviton spin wave. Background affine connection is attributed to exotic source of density field, $\eta_{\mu\nu}<\rho>$. The shifted expectation value, $\rho_{vac}\to\rho'$+$<\rho>_{rad}$, energy momentum tensor adds a constant term, $T_{\mu\nu}\to T'_{\mu\nu}$+$\eta_{\mu\nu}<\rho>_{rad}$, in scalar potential vacuum.

\section{De-sitter Spacetime}
Now we consider de-sitter spacetime under noncommutative manifold. In order to compare with the tensional geometry and the element disc, we get an adapt way to discuss how the behavior of a tensional geometry in curved structure. The geodesic line, Eq.($\hyperref[eq4]{4}$), can be written as
\begin{equation}\begin{split}\label{eq7}
d^{2}s' &= \eta_{ab}e^{a}_{~\mu}\otimes e^{b}_{~\nu}d\hat{x}^{\mu}\otimes d\hat{x}^{\nu}\\
            &= (\eta_{\mu\nu} + h_{\mu\nu} + \mathcal{B}_{\mu\nu})d\hat{x}^{\mu}\otimes d\hat{x}^{\nu}\\
            &= d^{2}s + (G^{S}_{\mu\nu} + G^{A}_{\mu\nu})dx^{\mu}dx^{\nu},\\
\end{split}\end{equation}
where
\begin{equation}\begin{split}\label{eq8}
d\hat{x}^{\mu}_{S}\otimes d\hat{x}^{\nu}_{S}&=\frac{1}{2}(d\hat{x}^{\mu}_{S}d\hat{x}^{\nu}_{S}+d\hat{x}^{\nu}_{S}d\hat{x}^{\mu}_{S})\\
d\hat{x}^{\mu}_{A}\otimes d\hat{x}^{\nu}_{A}&=\frac{1}{2}(d\hat{x}^{\mu}_{A}d\hat{x}^{\nu}_{A}+d\hat{x}^{\nu}_{A}d\hat{x}^{\mu}_{A})\\
d\hat{x}^{\mu}\otimes d\hat{x}^{\nu}&=\frac{1}{2}(d\hat{x}^{\mu}_{S}d\hat{x}^{\nu}_{S}+d\hat{x}^{\mu}_{A}d\hat{x}^{\nu}_{A})\\
E^{a}_{~\mu}\otimes E^{b}_{~\nu} &= \frac{1}{2}(E^{a}_{1\mu}E^{b}_{2\nu}+E^{a}_{2\nu}E^{b}_{1\mu}),\\
\end{split}\end{equation}
and $d^{2}s$ = $(\eta_{\mu\nu} + h_{\mu\nu})dx^{\mu}dx^{\nu}$ in $Riemann$ geometry. The graviton-like, $G^{A,S}_{\mu\nu}$, is a complete $\theta_{\mu\nu}$ deformed fields. Eq.($\hyperref[eq7]{7}$) gives a geometric connection between $Minkowski$ spacetime, $\eta_{\mu\nu}$, and $Riemann$ spacetime structure, $g_{\mu\nu}$. The graviton field takes a constraint on $vieblein$ tensor
\begin{equation}\begin{split}\label{eq9}
e^{0}_{~0} &= 1\\
e^{i}_{~0} &= e^{0}_{~i} = 0\\
e^{k}_{~i}e_{kj} &= - \delta_{ij} + h_{ij},
\end{split}\end{equation}
and the transformation between magnetic and non-magnetic frame is $d\hat{x}^{\mu}$ = $\hat{e}^{\mu}_{~\alpha}dx^{\alpha}$ with $vielbein$ field translation,
\begin{equation}\label{eq10}
\hat{e}^{a}_{~\mu} = \delta^{a}_{~\mu} + \theta^{a\lambda}n_{\lambda}\mathcal{K}\partial_{\mu}\rho,
\end{equation}
where $\rho$ is a massless $Nambu~Goldstone$ boson. In Ref.~\cite{Dvorkin:2007jp}, $Nambu~Goldstone$ boson can generate a temperature anisotropic effects on global universe without violating $Lorentz$ symmetry.

Imposing a condition on symmetric $\{a^{\dagger}_{m},a_{n}\}$ = 0 and anti-symmetric, $[c^{\dagger}_{m},c_{n}]$ = 0, commutator, null state includes a full degrees of freedom at elliptic framework. The null vector can be regarded as three unequal length orthogonal unit vectors, $n'^{\mu}_{m=1,2,3}$, and composed by the relation,
\begin{equation}\begin{split}\label{eq11}
n^{\mu}_{A} &= \sum_{m=1,2,3}a_{m}\hat{n}'^{\mu}_{m}\quad\{a^{\dagger}_{m},a_{n}\} = 0\\
n^{\mu}_{S} &= \sum_{m=1,2,3}c_{m}\hat{n}'^{\mu}_{m}\quad [c^{\dagger}_{m},c_{n}] = 0.
\end{split}\end{equation}
The conditions of $null$ vector are $n^{\mu}_{S}n_{S\mu}$ = 0, $n^{\mu}_{A}n_{A\mu}$ = 0, and the unit condition $\sum_{m}c^{2}_{m}$ = $\sum_{m}a^{2}_{m}$ = 1. Each factor operator, $a_{n}$ and $c_{n}$, in null vector, Eq.($\hyperref[eq11]{11}$), can compose a three independent coordinates. The ellipse universe framework is full in above restrictions.

We can regard graviton and non-abelian field are $Nambu~Goldstone$ field. In~\cite{Cho:1975sf}, it considers the spacetime in $d$ dimensions with $N = d$ internal symmetry under larger symmetry breaking. We consider $Pancar\check{e}$ transformation to generate a $Nambu~Goldstone$ field in the spacetime extension with extra 2 degrees in extending spacetime symmetry from the commutation relation in $null$ vectors. Eq.($\hyperref[eq1]{3}$) reads the translation between module and un-module frame, the $Nambu~Goldstone$ field is an exotic field in spacetime internal structure. The fluctuations in particle mass and an anomaly term are attributed to the extra degrees emerged in spacetime symmetry.

Splitting $vierbien$ field into two parts, $e^{a}_{~\mu}\hat{e}^{b}_{~\nu}\delta^{\mu}_{b}$, the degrees is amended to be 16 with the time dependent $Nambu~Goldstone$ boson. From Eq.($\hyperref[eq8]{8}$), the general product in this structure is
\begin{equation}\begin{split}\label{eq12}
d^{2}s &= (\eta_{\mu\nu} + h_{\mu\nu} + \mathcal{B}_{\mu\nu})d\hat{x}^{\mu}\otimes d\hat{x}^{\nu}\\
       &= \big{[}\eta_{\mu\nu} + \frac{1}{2}\big{(}G_{\mu\nu}(x,n_{S}) + G_{\mu\nu}(x,n_{A})\big{)}\big{]}dx^{\mu}dx^{\nu},
\end{split}\end{equation}
The abbreviations are $\frac{1}{2}G_{\mu\nu}(x,n_{S})$ = $G^{S}_{\mu\nu}$ + $\frac{1}{2}h_{\mu\nu}$ and $\frac{1}{2}G_{\mu\nu}(x,n_{A})$ = $G^{A}_{\mu\nu}$ + $\frac{1}{2}h_{\mu\nu}$. The rotated effects are devoted to graviton product with background field. The symmetric tensor is
\begin{equation}\begin{split}\label{eq13}
&G_{\mu\nu}(x,n_{S}) =\\
&\big{[}\theta_{\nu\lambda}n^{\lambda}_{S}\mathcal{K}\partial_{\mu}\rho + \theta_{\mu\lambda}n^{\lambda}_{S}\mathcal{K}\partial_{\nu}\rho + \theta_{\lambda\alpha}n^{\alpha}_{S}\theta^{\lambda\beta}n^{S}_{\beta}\mathcal{K}^{2}\partial_{\mu}\rho\partial_{\nu}\rho\big{]}\\
               &+ \big{[}h_{\mu\nu} + h_{\mu\alpha}\theta^{\alpha\beta}n^{S}_{\beta}\mathcal{K}\partial_{\nu}\rho + h_{\nu\alpha}\theta^{\alpha\beta}n^{S}_{\beta}\mathcal{K}\partial_{\mu}\rho\\
               &+ h_{\alpha\beta}\theta^{\alpha\eta}n^{S}_{\eta}\theta^{\beta\lambda}n^{S}_{\lambda}\mathcal{K}^{2}\partial_{\mu}\rho\partial_{\nu}\rho\big{]}\\
               &+ \big{[}B_{\mu\alpha}\theta^{\alpha\beta}n_{S\beta}\mathcal{K}\partial_{\nu}\rho + B_{\nu\alpha}\theta^{\alpha\beta}n^{S}_{\beta}\mathcal{K}\partial_{\mu}\rho\\
               &+ B_{\alpha\beta}\theta^{\alpha\eta}n^{S}_{\eta}\theta^{\beta\lambda}n^{S}_{\lambda}\mathcal{K}^{2}\partial_{\mu}\rho\partial_{\nu}\rho\big{]},
\end{split}\end{equation}
and the antisymmetric part is,
\begin{equation}\begin{split}\label{eq14}
&G_{\mu\nu}(x,n_{A}) =\\
&\big{[}\theta_{\nu\lambda}n^{\lambda}_{A}\mathcal{K}\partial_{\mu}\rho + \theta_{\mu\lambda}n^{\lambda}_{A}\mathcal{K}\partial_{\nu}\rho + \theta_{\lambda\alpha}n^{\alpha}_{A}\theta^{\lambda\beta}n^{A}_{\beta}\mathcal{K}^{2}\partial_{\mu}\rho\partial_{\nu}\rho\big{]}\\
               &+ \big{[}h_{\mu\nu} + h_{\mu\alpha}\theta^{\alpha\beta}n^{A}_{\beta}\mathcal{K}\partial_{\nu}\rho + h_{\nu\alpha}\theta^{\alpha\beta}n^{A}_{\beta}\mathcal{K}\partial_{\mu}\rho\\
               &+ h_{\alpha\beta}\theta^{\alpha\eta}n^{A}_{\eta}\theta^{\beta\lambda}n^{A}_{\lambda}\mathcal{K}^{2}\partial_{\mu}\rho\partial_{\nu}\rho\big{]}\\
               &+ \big{[}B_{\mu\alpha}\theta^{\alpha\beta}n_{A\beta}\mathcal{K}\partial_{\nu}\rho + B_{\nu\alpha}\theta^{\alpha\beta}n^{A}_{\beta}\mathcal{K}\partial_{\mu}\rho\\
               &+ B_{\alpha\beta}\theta^{\alpha\eta}n^{A}_{\eta}\theta^{\beta\lambda}n^{A}_{\lambda}\mathcal{K}^{2}\partial_{\mu}\rho\partial_{\nu}\rho\big{]},
\end{split}\end{equation}
the $null$ vector on the conditions $n'_{0}$ = 1, $\hat{n}'^{i}$=$\frac{\partial_{i}\rho}{\partial_{0}\rho}$, and the Klein-Gordon equation of $Nambu~Goldston$ boson is $\partial^{\mu}\rho\partial_{\mu}\rho$ = 0.

Eq.($\hyperref[eq13]{13}$) and Eq.($\hyperref[eq14]{14}$) can be rewritten into two parts, first is matter field and another is a term of interacting with background universe structure,
\begin{equation}\label{eq15}
h_{\mu\nu} + G^{S}_{\mu\nu} + G^{A}_{\mu\nu} = G^{m}_{\mu\nu} + G^{int}_{\mu\nu},
\end{equation}
in which $G^{m}_{\mu\nu}$ and $G^{int}_{\mu\nu}$ are
\begin{equation}\begin{split}\label{eq16}
G^{m}_{\mu\nu} &=h_{\mu\nu} + \frac{\mathcal{K}}{2}\bigg{[}h_{\mu\alpha}\theta^{\alpha\beta}n^{A}_{\beta}\partial_{\nu}\rho + h_{\nu\alpha}\theta^{\alpha\beta}n^{A}_{\beta}\partial_{\mu}\rho\bigg{]}\\
               &+\frac{\mathcal{K}^{2}}{2}\bigg{[}\theta_{\lambda\alpha}n^{\alpha}_{S}\theta^{\lambda\beta}n^{S}_{\beta} + B_{\alpha\beta}\theta^{\alpha\eta}n^{A}_{\eta}\theta^{\beta\lambda}n^{A}_{\lambda}\bigg{]}\partial_{\mu}\rho\partial_{\nu}\rho\\
G^{int}_{\mu\nu} &=\frac{\mathcal{K}}{2}\bigg{[}\theta_{\mu\lambda}n^{\lambda}_{S}\partial_{\nu}\rho + \theta_{\nu\lambda}n^{\lambda}_{S}\partial_{\mu}\rho\\
                 &+ B_{\mu\alpha}\theta^{\alpha\beta}n^{S}_{\beta}\partial_{\nu}\rho + B_{\nu\alpha}\theta^{\alpha\beta}n^{S}_{\beta}\partial_{\mu}\rho\\
                 &+ B_{\mu\alpha}\theta^{\alpha\beta}n^{A}_{\beta}\partial_{\nu}\rho + B_{\nu\alpha}\theta^{\alpha\beta}n^{A}_{\beta}\partial_{\mu}\rho\bigg{]}.
\end{split}\end{equation}
The term $\theta_{\nu\lambda}n^{\lambda}_{A}\mathcal{K}\partial_{\mu}\rho + \theta_{\mu\lambda}n^{\lambda}_{A}\mathcal{K}\partial_{\nu}\rho$ is absorbed into graviton gauge. Therefore, extracting graviton field
\begin{equation}\label{eq17}
h_{\mu\nu} + \frac{\mathcal{K}}{2}(h_{\mu\alpha}\theta^{\alpha\beta}n^{A}_{\beta}\partial_{\nu}\rho + h_{\nu\alpha}\theta^{\alpha\beta}n^{A}_{\beta}\mathcal{K}\partial_{\mu}\rho),
\end{equation}
photon mass term
\begin{equation}\label{eq18}
\frac{\mathcal{K}^{2}}{2}\theta_{\lambda\alpha}n^{\alpha}_{S}\theta^{\lambda\beta}n^{S}_{\beta}\partial_{\mu}\rho\partial_{\nu}\rho,
\end{equation}
and the term contains anomaly
\begin{equation}\label{eq19}
\frac{\mathcal{K}^{2}}{2}B_{\alpha\beta}\theta^{\alpha\eta}n^{A}_{\eta}\theta^{\beta\lambda}n^{A}_{\lambda}\partial_{\mu}\rho\partial_{\nu}\rho,
\end{equation}
in TT gauge $n^{\mu}_{A, S}h_{\mu\nu}$ = $n^{\mu}_{A, S}\mathcal{B}_{\mu\nu}$ = 0 and $h$ = $B$ = 0. The Lagrangian in matter, gauge field, and graviton field redefinition are complete included in the effects from module translation. In vacuum $\sqrt{-g}$ expansion, a part of graviton field, Eq.($\hyperref[eq16]{16}$), and background are absorbed into particle field normalization condition.

\section{Modify general relativistic}
In the study of universe structure, gravitation radiation is playing an important role as a membership of radiated particle field theory. As concerned in isotropic spacetime structure, radiation from any particle field does not associate with its position. The consequence is always a symmetric result. The structure injures the symmetry property in moving particle field radiation under anisotropic structure. Particle field is not only dependent on its situation, but it will be modified by adding a non-symmetric term as to the spacetime is not commutable.

Graviton radiation is a self-activity process, radiation process is induced by particle gaussian distribution. Particle is moving along to geodesic with finite energy region. The dispersion gets into energy loose due to the interaction with surroundings in classical mechanics. In quantum mechanics, it is devoted into uncertainty principle. The radiation in matter field, it is produced by the background affections. In gravitational case, the radiation is generated by the oscillation within itself.

From Eq.($\hyperref[eq17]{17}$), graviton field in TT gauge, the interaction between graviton spin field and background field will revise the equation of motion in observer. In the assumption of observer moving along z-axis and under graviton spinor field, the gravitational spin field is
\begin{equation}\label{eq20}
h_{\mu\nu} = \left(\begin{array}{@{}cccc@{}}
0 & 0 & 0 & 0\\
0 & \Delta_{+} & \Delta_{\times} & 0\\
0 & \Delta_{\times}& -\Delta_{+} & 0\\
0 & 0 & 0& 0
\end{array}\right).
\end{equation}
Hence, the $Cristoffel$ tensor can be written as
\begin{equation}\label{eq21}
\begin{split}
&\Gamma^{0}_{~00} = - (\vec{E}\cdot\hat{n})\mathcal{K}\partial_{0}(\dot{\rho}\phi)\\
&\Gamma^{1}_{~00} = - n_{0}\mathcal{K}\big{[}(\vec{E} + \vec{B}\times\hat{n})_{1}\partial_{0}(\dot{\rho}h_{+})\\
&\qquad\qquad+ (\vec{E} + \vec{B}\times\hat{n})_{2}\partial_{0}(\dot{\rho}h_{\times})\big{]}\\
&\Gamma^{2}_{~00} = - n_{0}\mathcal{K}\big{[}(\vec{E} + \vec{B}\times\hat{n})_{1}\partial_{0}(\dot{\rho}h_{\times})\\
&\qquad\qquad- (\vec{E} + \vec{B}\times\hat{n})_{2}\partial_{0}(\dot{\rho}h_{+})\big{]}\\
&\Gamma^{3}_{~00} = 0,\\
\end{split}
\end{equation}
and the equation of motion of particle coordinate is
\begin{equation}\label{eq22}
\begin{split}
&\frac{\delta^{2} x^{0}}{\delta^{2}\tau} = c^{2}(\vec{E}\cdot\hat{n})\mathcal{K}\partial_{0}(\dot{\rho}\phi)\\
&\frac{\delta^{2} x^{1}}{\delta^{2}\tau} = - c^{2}n_{0}\mathcal{K}\big{[}(\vec{E} + \vec{B}\times\hat{n})_{1}\partial_{0}(\dot{\rho}h_{+})\\
&\qquad\qquad+ (\vec{E} + \vec{B}\times\hat{n})_{2}\partial_{0}(\dot{\rho}h_{\times})\big{]}\\
&\frac{\delta^{2} x^{2}}{\delta^{2}\tau} = - c^{2}n_{0}\mathcal{K}\big{[}(\vec{E} + \vec{B}\times\hat{n})_{1}\partial_{0}(\dot{\rho}h_{\times})\\
&\qquad\qquad- (\vec{E} + \vec{B}\times\hat{n})_{2}\partial_{0}(\dot{\rho}h_{+})\big{]}\\
&\frac{\delta^{2} x^{3}}{\delta^{2}\tau} = 0.\\
\end{split}
\end{equation}
Abbreviating above form by extracting $x'^{1}$ and $x'^{2}$ coordinates, the oscillation in gravitational spinor background is
\begin{equation}\begin{split}\label{eq23}
&\frac{\partial^{2}}{\partial\tau^{2}}\begin{pmatrix}x^{1}\\x^{2}\end{pmatrix} =\\
&\mathcal{K}\begin{pmatrix}\partial_{0}(\dot{\rho}\Delta_{+}) & \partial_{0}(\dot{\rho}\Delta_{\times}) \\ \partial_{0}(\dot{\rho}\Delta_{\times})
&-\partial_{0}(\dot{\rho}\Delta_{+})\end{pmatrix}\begin{pmatrix}E^{1}+(\vec{B}\times\hat{n})^{1}\\ E^{2}+(\vec{B}\times\hat{n})^{2}\end{pmatrix}.\\
\end{split}\end{equation}
Therefore, in order to get rotation frequency, we have to introduce a matrix from diagonal Eq.($\hyperref[eq23]{23}$). The obtained rotation matrix $\mathcal{R}$,
\begin{equation}\label{eq24}
\mathcal{R} =
\begin{pmatrix}
\cos\theta_{h} & \sin\theta_{h}\\
-\sin\theta_{h} & \cos\theta_{h},\\
\end{pmatrix},
\end{equation}
connects with the gravitational field and background field under the basis of
\begin{equation}\label{eq25}
\begin{pmatrix}\ddot{x}^{1} \\ \ddot{x}^{2}\end{pmatrix}' = \mathcal{K}D\begin{pmatrix}E^{1}+(\vec{B}\times\hat{n})^{1} \\ -(E^{2}+(\vec{B}\times\hat{n})^{2})\end{pmatrix},
\end{equation}
and
\begin{equation}\label{eq26}
\begin{pmatrix}\ddot{x}^{1} \\ \ddot{x}^{2}\end{pmatrix}' = \begin{pmatrix}\cos\theta_{h} & \sin\theta_{h}\\-\sin\theta_{h} & \cos\theta_{h}\\ \end{pmatrix}\begin{pmatrix}\ddot{x}^{1}.\\ \ddot{x}^{2}\end{pmatrix}.
\end{equation}
The integration of each eigenvalues $D_{1}$ = $\int dl_{1}D$ and $D_{2}$ = $\int dl_{2}D$ in $dl_{1}$ and $dl_{2}$ are devoted into each departures in $x_{1}$ and $x_{2}$ coordinates. Hence, we get an energy displacement $D_{1}$ and $D_{2}$ from the force in the devotion of universe rotation.

The universe rotation attributes to the eigenvalues $D_{1}$ and $D_{2}$. It implies that the observed universe is not a perfect spherical symmetric. The potential along $x^{1}$ and $x^{2}$ axis are
\begin{equation}\begin{split}\label{eq27}
\phi'_{1} &= \mathcal{K}D_{1}\big{(}(\vec{B}\times\hat{n}) + \vec{E}\big{)}^{1}\\
\phi'_{2} &= -\mathcal{K}D_{2}\big{(}(\vec{B}\times\hat{n}) + \vec{E}\big{)}^{2}.\\
\end{split}\end{equation}
In the equal time, $t$, coordinate, observer is not standing on particle inertial frame, but a distance from the central incident point. The according transformation is consistent with the relation between different frame. The complete transformation from a rotated object to observer is
\begin{equation}\label{eq28}
\frac{d\mathcal{U}^{\mu}}{dt} + \Gamma^{\mu}_{\alpha\beta}\mathcal{U}^{\alpha}\mathcal{U}^{\beta} = - \frac{d^{2}t/d^{2}\tau}{(dt/d\tau)^{2}}\mathcal{U}^{\mu},
\end{equation}
where $\mathcal{U}^{\mu}$ = $\frac{dx^{\mu}}{dt}$. From the proper time to equal time, observer is standing on the different frame with a velocity correction between each others. The right handed side is a term dependent of the relative velocity in observer. In front of the right handed side in Eq.($\hyperref[eq28]{28}$), includes the transformation between proper time, $\tau$, and equal time, $t$. The minus sign is to satisfy the energy momentum conservation law in full system. The observed result is still in a rotated objects. The different time translation is similar to rotate or boost an elliptic universe in an angle or a shift.

For the macroscopic system, the solar system, each planet is moving on an elliptic trajectory by $Newton$ force. On the contrast, in field distribution, particle itself suffers an elliptic force in the field redefinition under these background surroundings. Thus, we get the modified $Newton$ potential in an oscillated and rotated universe situation.

\section{Inflation and Oscillation induce anomaly}
Background field is an uniformed field in whole of the universe. Each particle interacts with universe background via the transformation from non-background frame to background field site. We assume that surroundings is merged in elliptic null frame, however anomaly cancellation is manifest on this framework automatically.

Fermion chiral gauge transformation is not conserved in fermion field with considering a phase in its geodesic line. Theoretically, vacuum has to contain an extra phase factor with $\gamma_{5}$ generator. Ref.~\cite{Fujikawa:1979ay} imposes a cutting rule for nonphysical state in $Jacobian$ factor determination. In the module field extension, there no nonphysical field is produced in vacuum. All fields in module translation are well defined. In supersymmetry model~\cite{DeRydt:2008hw}, it shows that $Green~Schwarz$ anomaly cancelation is a source to generate a relation between magnetic and electric duality. Magnetic field and electric field are a dual fields, and their product injure the dual symmetry in $Maxwell$ action. Anomaly is the same form as $Chern~Simons$ term in three dimensions or higher odd dimensions. It is a term generating baryogenesis and leptogenesis symmetry under hypersurface out of the thermal equilibrium.

In this model, we extract the most important part into consideration of neutrino, photon mass and anomaly phenomena. The other terms, affine field, are devoted into field normalization. As the reason, particle field is expanded by considering an energy stored effects. It should be perturbative expanded in $\theta_{\mu\nu}$ next leading order. In noncommutative spacetime, fermion field automatically survives in the background field ambience. Particle always survives in the background frame if observer locates near the region from background field induced curved spacetime. Particle's equation of motion is the same as the coordinate in $\hat{x}^{\mu}$ frame. By contrary, if we are far away from the region, we similar reside in the free falling coordinate. Photon is a null field, it does not contain the other degrees of freedom to generate its mass in vacuum except $Lorentz$ is not exactly symmetry. If we consider photon field on dual background, there extra degrees of freedom is associated with those and the $null$ vector degrees of freedom.

In Eq.($\hyperref[eq16]{16}$), anomaly is extracted in Eq.($\hyperref[eq19]{19}$) contains an uniformed bosonic gas in global universe. The symmetric property in geometry, we get an exact term depending on the antisymmetric $null$ vector, $n^{A}_{\mu}$, and the antisymmetric gravitational field, $B_{\mu\nu}$. The extracted anomaly from Eq.($\hyperref[eq19]{19}$) is
\begin{equation}\label{eq29}
\mathcal{K}^{2}\big{[}2(\vec{H}\cdot\vec{n}^{A})n^{A}_{0} - \vec{A}\cdot(\vec{n}^{A}\times\vec{n}^{A})\big{]}\epsilon_{0ijk}\theta^{0i}\theta^{jk}\partial_{\mu}\rho\partial_{\nu}\rho,
\end{equation}
where $A^{i}$ = $B^{0i}$, and $H^{i}$ = $\frac{1}{2}\epsilon^{ijk}B_{jk}$. It is associated with the module field, $\rho$, and dependent with its derivative $\partial_{0}\rho$ and $\partial_{\vec{x}}\rho$. The kinetic momentum $\mathcal{K}$ is a shift from particle horizon field. The term, $\epsilon_{0ijk}\theta^{0i}(x)\theta^{jk}(x)$, is a $Lorentz$ conserved term in considering an inner structure in $\theta_{\mu\nu}(x)$. The general $U(1)$ anomaly includes $\bf{E}\cdot\bf{B}$ combines an asymmetric property between electric field and magnetic field. In Eq.($\hyperref[eq30]{29}$), the form in front of the anomaly term is a second order devotion in $Jacobian$ factor. It induces a phase transition from the inertial frame to magnetic frame. In the viewpoint of phase transition, anomaly effective lagrangian is generated from an unstable vacuum, it is a possible parity violation from topological effect.

In the convenience, the combination between symmetric and anti-symmetric metric devotions without considering the $\theta_{\mu\nu}$ noncommutative bosonic gas in model building. The complete lagrangian in fermion and gauge field
\begin{equation}\begin{split}\label{eq30}
&\mathcal{L} =\\
&\int d^{4}x\bigg\{\frac{i}{2}E_{a}^{~\mu}\bar{\psi}\gamma^{a}\overleftrightarrow{D}_{\mu}\psi -m\bar{\psi}\psi + \frac{w_{bcA}}{8}\bar{\psi}\{\gamma^{A}, \sigma^{bc}\}\psi\\
&+ \phi(x)\frac{e^{2}}{16\pi^{2}}\epsilon_{\mu\nu\alpha\beta}\theta^{\mu\nu}(x)\theta^{\beta\alpha}(x) - \frac{1}{4}F_{\mu\nu}F^{\mu\nu}\\
&+ \frac{\rho_{0}\rho_{\vec{x}}}{2\Lambda^{2}_{C}}\theta^{global}_{\lambda\alpha}(x)n^{\alpha}_{S}\theta^{\lambda\beta}_{global}(x)n^{S}_{\beta}\bigg{\}} + O(\theta^{3}),\\
\end{split}\end{equation}
and the gauge potential
\begin{equation}\label{eq31}
\theta_{\mu\nu}(x) = \partial_{\mu}\mathcal{A}^{global}_{\nu} - \partial_{\nu}\mathcal{A}^{global}_{\mu},
\end{equation}
is builded on non-local $U(1)$ gauge potential $\mathcal{A}^{global}_{\mu}$, where $\theta^{global}_{\mu\nu}(x)$ is a dimension two tensor by absorbing $\mathcal{K}^{2}$ into its gauge potential and the derivative operator and $\phi(x)$ is axion field. The total $vieblein$ field, $E^{a}_{~\mu}$ = $e^{a}_{~\lambda}\hat{e}^{b}_{~\mu}\delta^{\lambda}_{b}$, transfers curved frame to free falling frame and its inverse $vieblein$ field can be expressed as $E^{~a}_{\mu}$ = $g_{\mu b}\eta^{b\lambda}E^{a}_{~\lambda}$. The vector potential and axial potential are regarded as $V_{\mu}$ = $V_{a}E^{a}_{~\mu}$ and $A_{\mu}$ = $A_{a}E^{a}_{~\mu}$ respectively.

There are an embedded degrees of freedom produced by three dimensions background location and null vector direction. Particle current, whether vector or axial current, is conserved in the noncommutative background with an uniformed $null$ state. The conservation law are
\begin{equation}\begin{split}\label{eq32}
&\mathcal{D}_{\mu}\mathcal{J}^{a\mu} + e\epsilon^{abc}A^{b}_{\mu}\mathcal{J}^{c\mu} = 0\\
&\mathcal{D}_{\mu}\mathcal{J}^{a\mu}_{5} +e\epsilon^{abc}A^{b}_{\mu}\mathcal{J}^{c\mu} = 2im\mathcal{P}^{a} + \mathbf{A},\\
\end{split}\end{equation}
in which $\mathbf{A}$ is U(1) anomaly term, the index a is 0, 1, 2, 3, and 0 denotes abelian group. In non-abelian theory, $ABJ$ anomaly is generated by U(1) gauge with axial symmetry. There a condition is restricted into model building, $Tr\{C_{a},\{C_{b},C_{c}\}\}\}$ = 0, where $C_{a,b,c}$ is non-abelien generator. In the commutative field theory, it is produced by axial current coupled to two vector triangle diagrams.

\section{Neutrino Mass and Neutrino Oscillation by Graviton Radiation}
Violating $Lorentz$ invariance generates a modification of dispersion relation~\cite{Stecker:2004vm}. In which the general expression of modifying particle energy spectrum induces a rotation between different generations of particle energy mixing effects. $Lorentz$ violated theory with $CPT$ conservation gives a clue in neutrino oscillation~\cite{Dighe:2008bu}. $CPT$ conservation implies particle is survived in a statable situation. Thermal equilibrium is preserved in the universe from a statistic condition. However, $CPTV$ theorem implies $Lorentz$ invariance is not conserved~\cite{Greenberg:2002uu}, particle thermal equilibrium slightly amends the accordance to tinily phenomena of universe inflation.

In dark matter model, the field in universe vacuum produces a mass term to massless particle~\cite{Zee:1985id, Babu:1988ki,Hung:2006ap,Barbieri:2005gj} by considering pesudo $Nambu~Goldstone$ potential. It generates an oscillation effect between different generations. In the collider experiment on nucleon target, neutrino oscillation effects between $K$-shell is frequently detected~\cite{Kleinert:2008ps}.

We consider a background field fluctuations in universe vacuum. Inverse $vielbein$ transformation preserves the thermal equilibrium in dimensionless phase transition. The affection does not generate an influence of universe acceleration, but devote on the product of $Nambu~Goldstone$ particle with background field. The exotic interaction with background generates an induced correlation in considering a non-traceless geometry and background field in universe. Exchanging the basis from universe vacuum $|0>$ to background eigenstate $|\Omega>$, the radius imposes a phase factor $e^{i\frac{\pi}{4}}$ in $r$ with a freezed universe. There energy store into a volume by integrating out the global angle within the deviated debris $\delta r$. Hence, the energy in generating particle mass can be extracted from the induced horizon, $\delta r$, of particle field.

The energy momentum tensor from module translation is,
\begin{equation}\begin{split}\label{eq33}
T'^{\mu\nu} = T^{\mu\nu} + T^{\mu\lambda}\theta^{\lambda\eta}n_{\eta}\mathcal{K}\partial_{\nu}\rho + T^{\nu\lambda}\theta^{\lambda\eta}n_{\eta}\mathcal{K}\partial_{\mu}\rho,\\
\end{split}\end{equation}
and harmonic oscillation condition, $\eta^{\alpha\beta}\Gamma^{\mu}_{~\alpha\beta}$ = 0. Thus, graviton spin field can be written as
\begin{equation}\label{eq34}
e'^{\mu\nu} = e^{\mu\nu} + e^{\mu\lambda}\theta^{\lambda\eta}n_{\eta}\mathcal{K}\partial_{\nu}\rho + e^{\nu\lambda}\theta^{\lambda\eta}n_{\eta}\mathcal{K}\partial_{\mu}\rho,
\end{equation}
where
\begin{equation}\label{eq35}
k^{2}e^{\mu\nu} = T^{\mu\nu} - \frac{1}{2}\eta^{\mu\nu}T.
\end{equation}
The translation to free falling frame, we get the energy-momentum tensor a primed form
\begin{equation}\begin{split}\label{eq36}
T'^{\mu\nu} &= T^{\mu\nu} + \big{[}T^{\mu\lambda}\theta_{\lambda\eta}n^{\eta}\mathcal{K}\partial^{\nu}\rho + (\mu\leftrightarrow\nu)\big{]}\\
                    &- \frac{1}{2}T\big{[}\theta^{\mu\eta}n_{\eta}\mathcal{K}\partial^{\nu}\rho + (\mu\leftrightarrow\nu)]\\
                    &- \bigg{\{}\eta^{\gamma\nu}[T_{\gamma\lambda}\theta^{\lambda\eta}n_{\eta}\mathcal{K}\partial^{\mu}\rho - \frac{1}{2}T\theta^{\gamma\eta}n_{\eta}\mathcal{K}\partial^{\mu}\rho\big{]}\\
                    & + (\mu\leftrightarrow\nu)\bigg{\}}.\\
\end{split}\end{equation}

The general fermion energy momentum tensor,
\begin{equation}\label{eq37}
T^{\mu\nu}_{\psi} = \frac{i}{2}\bar{\psi}\gamma^{\{\mu,}\partial^{\nu\}}\psi - \eta^{\mu\nu}(i\bar{\psi}\not{\partial}\psi - m_{\psi}\bar{\psi}\psi),
\end{equation}
preserves fermion is a on shell particle. The external particle horizon is induced from universe basis and the rotated generation. By extracting the term
\begin{equation}\label{eq38}
-\eta_{\gamma\nu}T^{\gamma}_{~\lambda}\theta^{\lambda\eta}n_{\eta}\mathcal{K}\partial_{\mu}\rho + (\mu\leftrightarrow\nu),
\end{equation}
from Eq.($\hyperref[eq36]{36}$) and the kinetic term from Eq.($\hyperref[eq37]{37}$) and putting fermion energy momentum tensor, Eq.($\hyperref[eq37]{37}$), into Eq.($\hyperref[eq38]{38}$), it obtains that
\begin{equation}\label{eq39}
-(J^{\nu}_{\psi}p_{\lambda} + p^{\nu}j^{\psi}_{\lambda})\theta^{\lambda\eta}n_{\eta}\mathcal{K}\partial^{\mu}\rho + (\mu\leftrightarrow\nu),
\end{equation}
in which the fermion current interaction between background debris is
\begin{equation}\label{eq40}
J_{\psi\lambda}\theta^{\lambda\eta} = 0,\quad J^{\nu}_{\psi}\partial^{\mu}\rho + (\mu\leftrightarrow\nu) = 2\eta^{\mu\nu}\bar{\psi}\not\partial\rho\psi,
\end{equation}
where $J_{\psi\mu}$ = $\bar{\psi}\gamma_{\mu}\psi$. By absorbing $G_{BG}$ into particle field horizon, we choose the ground basis as $|\Omega>$. The equation, $i\not{\partial}\rho|\Omega>$ = -~$\frac{i}{2}\delta m_{\rho}|\Omega>$, is taking into account which the exotic mass term devotes from particle horizon effects.

Therefore, neutrino mass is generated from the universe ground state with the radius $r'\to re^{i\frac{\pi}{4}}$. The mass term is corrected by integrating out the shifted radius $\delta r$. However, the braket expectation value is coming from the integration of the remains from changed basis,
\begin{equation}\begin{split}\label{eq41}
<0|G_{BG}&(\delta r)|0>|_{r'\to re^{i\frac{\pi}{4}}, \delta t\to 0}\\
                      &\to -i\big{(}Re(\delta m_{\rho}) + Im(\delta m_{\rho})\big{)}<\Omega|\Omega>,\\
\end{split}\end{equation}
where $|0>$ is free falling vacuum state. The image mass part can be absorbed into $Yukawa$ coupling inducing a possible $CPV$ decay process and the real part attributed to particle mass correction. Where $G_{BG}$ is the determination of background field metric, particle total mass term is $\eta_{\mu\nu}(m_{\psi} + \delta m_{\psi})\bar{\psi}\psi$. The neutrino field, $m_{\psi}$ = 0, its mass term cannot be generated by $Yukawa$ coupling from spontaneous breaking. The module field is an extra source to generate particle mass, $\delta m_{\psi}$. Neutrino and photon will generate its mass from universe vacuum fluctuations with module translation. The fermion action is rewritten as
\begin{equation}\label{eq42}
\mathcal{L}_{\psi} = i\bar{\psi}\gamma^{\mu}\partial_{\mu}\psi - (m_{\psi} + \delta m_{\rho})\bar{\psi}\psi.
\end{equation}

However, neutrino mass generates from the interaction with module field and stored background energy from rotated outer horizon field. Furthermore, the one loop photon correction, fermion mass is fluctuated around its mass eigenvalue. The divergent effect absorbes into vacuum electric energy from the source of the charge distribution.

\section{Photon Mass Production}
In the background metric, vacuum energy interacts with particle field and contributes on the horizon. We have discussed that the horizon effects induces fermion mass under above section. The contribution on the massless gauge bason, photon, involves the same devotion on particle mass production. In this section, we discuss the photon mass production from the assumed symmetric $null$ vector in geometry extension, Eq.($\hyperref[eq18]{18}$).

Under the geometry determination, $G^{m}_{\mu\nu}$ + $G^{int.}_{\mu\nu}$, the nonlocal background strength tensor constructs an inner structure in field normalization $\theta_{\mu\nu}\to\theta_{\mu\nu}(x)$. We get a term, $\frac{1}{2}\theta_{\lambda\alpha}n^{\alpha}\theta^{\lambda\beta}n_{\beta}$, to depict a background field energy distribution into universe. Due to particle entropy is not an uniformed distribution in global universe. The induced mass from $null$ vector and $Nambu~Goldstone$ boson $\rho$ depends on particle field location. The photon field action adds a mass term into Lagrangian
\begin{equation}\label{eq43}
\mathcal{L}_{ph} = -\frac{1}{2}\big{[}\partial_{\mu}A^{\dagger}_{\nu}\partial^{\mu}A^{\nu} + \mu^{2}A^{2}_{global}\big{]},
\end{equation}
with the mass $\mu^{2}$ = $\frac{\rho_{0}\rho_{\vec{x}}}{\Lambda^{2}_{C}}k^{2}_{n}$, where $k^{2}_{n}$ = $k^{0}n_{0}$ - $\vec{k}\cdot\vec{n}\cos\theta_{k}$, and $\theta_{k}$ is the angle between $null$ vector and photon kinetic momentum. It associates with photon kinetic momentum belongs to the null direction.

As we consider an azimuth orientation, the mass term can be integrated out in standing on photon inertial frame. It contains a background field spectrum and a preferred direction. The gauge potential $A^{\mu}$ still satisfies the gauge condition in null direction, $\dot{\rho}n^{i}//\partial_{i}\rho$. Therefore, $A^{\mu}\to A^{\mu} + \partial^{\mu}\rho$ should add a mass term in action with $null$ gauge condition. In order to depict a mass term from gauge boson, the null gauge freedom on vector field has to be imposed, $n_{\mu}A^{\mu}$ = $\partial_{\mu}\rho A^{\mu}$ = 0. In which, $\partial_{n}$ = $n^{\mu}\partial_{\mu}$, denotes photon kinetic term along the direction of null vector. Which the polarization depends on the selected gauge, it will not be modified by considering a null vector in the background universe. Photon propagator is rewritten by adding a mass term
\begin{equation}\begin{split}\label{eq44}
<0|T(\mathcal{A}_{\mu}(k)&\mathcal{A}_{\nu}(k))|0> =\\
                         &\frac{-i}{q^{2}-\mu^{2}}\bigg{[}g_{\mu\nu} - \frac{n_{\mu}k_{\nu} + n_{\nu}k_{\mu}}{(k\cdot n)^{2}} + n^{2}\frac{k_{\mu}k_{\nu}}{(k\cdot n)^{2}}\bigg{]},\\
\end{split}\end{equation}
where $k_{n}$ is photon momentum along to null direction. The kinetic momentum $k_{n}$ is not the same for the extended degrees of freedom in a short length.

For an equal length of each orthogonal directions, the generated phenomenon is a spherically symmetric universe. On the other hand, each null state in a fully expanded degrees of freedom, parity is automatically violated. In which universe background eigenstate, we have to search a parity violation in virtual gauge boson propagator in considering a signal gauge boson propagator, whether loop diagrams or tree level processes from high energy collider experiments. The most general parity violation effects in U(1) model is resulting from $Chern~Simons$ density~\cite{Jackiw:1984gg}, $\Omega^{\mu} = \frac{1}{8\pi^{2}}tr\epsilon^{\mu\alpha\beta\gamma}(A_{\alpha}\partial_{\beta}A_{\gamma} + \frac{2}{3}A_{\alpha}A_{\beta}A_{\gamma})$. $Chern~Simons$ is an anomaly term in topological description $\partial_{\mu}\Omega^{\mu}$. The triple photon coupling from $Chern~Simons$ term vanishes on isotropic universe, but that will be appeared in the parity violated background, such as $CPT$ violation or noncommutative theory, etc.. In the $CPT$ violated theory, $Chern~Simons$ term is written as $\frac{1}{2}(k_{AF})^{\kappa}\epsilon_{\kappa\lambda\mu\nu}A^{\lambda}F^{\mu\nu}$~\cite{Colladay:1996iz, Kostelecky:2003fs}. Parity is inherent violated in each unique cell, however, the induced phenomenon of violating $Lorentz$ boost and rotation is manifestly\footnote{Noncommutative $Chern~Simons$, $\frac{1}{2}k^{\alpha}\epsilon_{\alpha\beta\mu\nu}tr[\hat{A}^{\beta}\star\partial^{\mu}\hat{A}^{\nu} + \frac{2}{3}\hat{A}^{\beta}\star\hat{A}^{\mu}\star\hat{A}^{\nu}]$~\cite{Chu:2000bz}, depends on the $\star$ product.}.

We depict a elliptic universe under isotropic background, the generated pressure will induce a non-shperical profile. Entropy is playing an important role to model a cosmology framework or particle theory. In different species universes, the entanglement between different eigenstate overlap rotates each other to a mixed situation. Thus, by considering an objects in elliptic ball, different field distribution couples to each other from the interaction in standard model. In this model, it puts an universe background under a rotated surroundings, and considers a tensional field, modulation field, into particle string. Also, it will generate an anomaly, Eq.($\hyperref[eq30]{30}$), in the spherical symmetry, and induce a rotation effects, Eq.($\hyperref[eq23]{23}$), in isotropic universe background.

Due to entropy distribution is not isotropic, U(1) should be parity violated. By producing momentum $p^{\mu}$ to a vector, as the momentum conserves in $Lorentz$ transformation, the parity violated phenomena cannot be produced. On the other hand, if the four vectors momentum is universal constant, the other covariant vector will be a parity violated source, such as $Chern~Simons$ term. In the module relativistic theory, those terms can be produced in assuming an unequal unit vector in euclidian spacetime structure.

The mass depends on the difference in particle field $\rho$. The interaction of background interference, it obtains a revised distribution in particle wave function and the translation from a surroundings to another by phase transition. In gauge boson field, it generates a mass term associated with background $Maxwell$ Lagrange. A global strength tensor $\theta_{\mu\nu}(x)$ can be generally written as $\partial_{\mu}A^{global}_{\nu} - \partial_{\nu}A^{global}_{\mu}$ in Eq.($\hyperref[eq32]{32}$). In which, the background potential $A^{global}_{\mu}$ is a massive field in the eigenstate $|\Omega>$ with the modified $Maxwell$ action
\begin{equation}\label{eq45}
-\frac{1}{4}F^{\mu\nu}F_{\mu\nu} + \frac{\rho_{0}\rho_{\vec{x}}}{2\Lambda^{2}_{C}}\theta_{\alpha\mu}(x)n^{\mu}\theta^{\alpha\nu}(x)n_{\nu},
\end{equation}
the second term includes a dimension two $Maxwell$ field and depends on the generated photon mass. The spontaneous breaking of scalar potential to vacuum with the deviation of fermion mass. The massless gauge boson in module background can generate a mass term in a shadow of the map from Einstein to curved magnetic coordinates.

\section{Conclusion}
Quantized bosonic gas in noncommutative deformation generates with the debris under universe background. However, we use the noncommutative relation to impose a deformed spacetime in module translation. $Nambu~Goldstone$ boson plays an important role to generate curved universe. Thus, we find that the background universe is rotated due to null vector couples to background strength field of violating $Lorentz$ symmetry. Assuming background field is a global constant and the null vector exists in a non-spherical spacetime. The important consequence is that particle field is self rotated in graviton spinor surroundings with noncommutative module extension and $Lorentz$ violated surroundings under noncommutative bosonic gas ambience without considering $Weyl~Moyal$ product. Rotation effects induces a non-spherical effects, e.g. $ABJ$ anomaly cancellation. Considering a non-traceless background stored energy via rotating radius in $\frac{\pi}{4}$ degrees with the freezed universe, we get an added particle mass fluctuations from stored energy by integrating out the global angle and the displacement $\delta r$. However, the universe background devotes on fermion mass term, e.g. generating neutrino and photon mass.

\acknowledgments{We will thank Yi-Yang for useful discuss. This work is supported in part by the National Science Council of R.O.C. under contact : NSC-95-2112-M-007-059-MY3 and National Tsing Hua University under contact : 97N2309F1.}


\begin{thebibliography}{99}
\bibitem{Seiberg:1999vs}
  N.~Seiberg and E.~Witten,
  JHEP {\bf 9909}, 032 (1999)
  [arXiv:hep-th/9908142];
  N.~Seiberg, L.~Susskind and N.~Toumbas,
  JHEP {\bf 0006}, 021 (2000)
  [arXiv:hep-th/0005040];
A. Abouelsaood, C. G. Callan, C. R. Nappi and S. A. Yost, Nucl. Phys. {\bf B280}, 599 (1987);\\

\bibitem{Martin:2002nr}
  C.~P.~Martin,
  Nucl.\ Phys.\  B {\bf 652}, 72 (2003)
  [arXiv:hep-th/0211164];

\bibitem{Carroll:2001ws}
  S.~M.~Carroll, J.~A.~Harvey, V.~A.~Kostelecky, C.~D.~Lane and T.~Okamoto,
  Phys.\ Rev.\ Lett.\  {\bf 87}, 141601 (2001)
  [arXiv:hep-th/0105082].

\bibitem{Banerjee:2005zt}
  R.~Banerjee,
  Eur.\ Phys.\ J.\  C {\bf 47}, 541 (2006)
  [arXiv:hep-th/0508224].

\bibitem{Chen:2008mh}
  C.~Y.~Chen,
  arXiv:0808.2848 [hep-ph].

\bibitem{Barosi:2008gx}
  L.~Barosi, F.~A.~Brito and A.~R.~Queiroz,
  JCAP {\bf 0804}, 005 (2008)
  [arXiv:0801.0810 [hep-th]].

\bibitem{Bemfica:2005pz}
  F.~S.~Bemfica and H.~O.~Girotti,
  J.\ Phys.\ A  {\bf 38}, L539 (2005)
  [arXiv:quant-ph/0506191].

\bibitem{Filk:1996dm}
  T.~Filk,
  Phys.\ Lett.\  B {\bf 376} (1996) 53.

\bibitem{Greenberg:2005jq}
  O.~W.~Greenberg,
  Phys.\ Rev.\  D {\bf 73}, 045014 (2006)
  [arXiv:hep-th/0508057].

\bibitem{Colatto:2005fy}
  L.~P.~Colatto, A.~L.~A.~Penna and W.~C.~Santos,
  Phys.\ Rev.\  D {\bf 73}, 105007 (2006)
  [arXiv:hep-th/0512266].

\bibitem{Banerjee:2005zq}
  R.~Banerjee and K.~Kumar,
  Phys.\ Rev.\  D {\bf 72}, 085012 (2005)
  [arXiv:hep-th/0505245].

\bibitem{WQFT} Weiberg, The Quantum Theory of Fields 1996;\\

\bibitem{Bertolami:2002eq}
  O.~Bertolami and L.~Guisado,
  Phys.\ Rev.\  D {\bf 67}, 025001 (2003)
  [arXiv:gr-qc/0207124].
  
\bibitem{Bertolami:2005hz}
  O.~Bertolami,
  Mod.\ Phys.\ Lett.\  A {\bf 20}, 1359 (2005)
  [arXiv:gr-qc/0501058].

\bibitem{Buric:2006di}
  M.~Buric, T.~Grammatikopoulos, J.~Madore and G.~Zoupanos,
  JHEP {\bf 0604}, 054 (2006)
  [arXiv:hep-th/0603044].

\bibitem{Rivelles:2002ez}
  V.~O.~Rivelles,
  Phys.\ Lett.\  B {\bf 558}, 191 (2003)
  [arXiv:hep-th/0212262].

\bibitem{Brandt:2006ua}
  F.~T.~Brandt and M.~R.~Elias-Filho,
  Phys.\ Rev.\  D {\bf 74}, 067704 (2006)v
  [arXiv:hep-th/0609106].

\bibitem{Bluhm:2004ep}
  R.~Bluhm and V.~A.~Kostelecky,
  Phys.\ Rev.\  D {\bf 71}, 065008 (2005)
  [arXiv:hep-th/0412320];
  R.~Bluhm,
  arXiv:0801.1711 [gr-qc].

\bibitem{Cohen:2006ky}
  A.~G.~Cohen and S.~L.~Glashow,
  Phys.\ Rev.\ Lett.\  {\bf 97}, 021601 (2006)
  [arXiv:hep-ph/0601236].
  G.~W.~Gibbons, J.~Gomis and C.~N.~Pope,
  Phys.\ Rev.\  D {\bf 76}, 081701 (2007)
  [arXiv:0707.2174 [hep-th]].
  E.~Alvarez and R.~Vidal,
  Phys.\ Rev.\  D {\bf 77}, 127702 (2008)
  [arXiv:0803.1949 [hep-th]].
  W.~Muck,
  arXiv:0806.0737 [hep-th].
  M.~M.~Sheikh-Jabbari and A.~Tureanu,
  arXiv:0806.3699 [hep-th].

\bibitem{Kostelecky:1988zi}
  V.~A.~Kostelecky and S.~Samuel,
  Phys.\ Rev.\  D {\bf 39}, 683 (1989);

\bibitem{Kostelecky:2003fs}
  V.~A.~Kostelecky,
  Phys.\ Rev.\  D {\bf 69}, 105009 (2004)
  [arXiv:hep-th/0312310];

\bibitem{Myers:2003fd}
  R.~C.~Myers and M.~Pospelov,
  Phys.\ Rev.\ Lett.\  {\bf 90}, 211601 (2003)
  [arXiv:hep-ph/0301124];
  
\bibitem{Hebecker:2003iw}
  A.~Hebecker and C.~Wetterich,
  Phys.\ Lett.\  B {\bf 574}, 269 (2003)
  [arXiv:hep-th/0307109].
  C.~Wetterich,
  Phys.\ Rev.\  D {\bf 70}, 105004 (2004)
  [arXiv:hep-th/0307145].

\bibitem{Mashhoon:2008vr}
  B.~Mashhoon,
  arXiv:0805.2926 [gr-qc].

\bibitem{Bastrukov:2008fh}
  S.~I.~Bastrukov, I.~V.~Molodtsova, S.~Misicu, H.~K.~Chang and D.~V.~Podgainy,
  arXiv:0804.0068 [nucl-th];

\bibitem{Dvorkin:2007jp}
  C.~Dvorkin, H.~V.~Peiris and W.~Hu,
  Phys.\ Rev.\  D {\bf 77}, 063008 (2008)
  [arXiv:0711.2321 [astro-ph]].

\bibitem{Cho:1975sf}
  Y.~M.~Cho and P.~G.~O.~Freund,
  Phys.\ Rev.\  D {\bf 12}, 1711 (1975).

\bibitem{Fujikawa:1979ay}
  K.~Fujikawa,
  Phys.\ Rev.\ Lett.\  {\bf 42}, 1195 (1979).

\bibitem{DeRydt:2008hw}
  J.~De Rydt, T.~T.~Schmidt, M.~Trigiante, A.~Van Proeyen and M.~Zagermann,
  arXiv:0808.2130 [hep-th].

\bibitem{Stecker:2004vm}
  F.~W.~Stecker,
  Int.\ J.\ Mod.\ Phys.\  A {\bf 20}, 3139 (2005)
  [arXiv:astro-ph/0409731];
  S.~R.~Coleman and S.~L.~Glashow,
  Phys.\ Rev.\  D {\bf 59}, 116008 (1999)
  [arXiv:hep-ph/9812418];

\bibitem{Dighe:2008bu}
  A.~Dighe and S.~Ray,
  arXiv:0802.0121 [hep-ph];
 V.~A.~Kostelecky,
  arXiv:hep-ph/0403088;
V.~A.~Kostelecky and M.~Mewes,
  Phys.\ Rev.\  D {\bf 70}, 031902 (2004)
  [arXiv:hep-ph/0308300];

\bibitem{Greenberg:2002uu}
  O.~W.~Greenberg,
  Phys.\ Rev.\ Lett.\  {\bf 89} (2002) 231602
  [arXiv:hep-ph/0201258];

\bibitem{Zee:1985id}
  A.~Zee,
  Nucl.\ Phys.\  B {\bf 264}, 99 (1986).

\bibitem{Babu:1988ki}
  K.~S.~Babu,
  Phys.\ Lett.\  B {\bf 203}, 132 (1988).

\bibitem{Hung:2006ap}
  P.~Q.~Hung,
  Phys.\ Lett.\  B {\bf 649}, 275 (2007)
  [arXiv:hep-ph/0612004].
  
\bibitem{Barbieri:2005gj}
  R.~Barbieri, L.~J.~Hall, S.~J.~Oliver and A.~Strumia,
  Phys.\ Lett.\  B {\bf 625}, 189 (2005)
  [arXiv:hep-ph/0505124].

\bibitem{Kleinert:2008ps}
  H.~Kleinert and P.~Kienle,
  arXiv:0803.2938 [nucl-th];

\bibitem{Jackiw:1984gg}
C.~Soo,
  Phys.\ Rev.\  D {\bf 59}, 045006 (1999)
  [arXiv:hep-th/9805090];
  H.~Falomir, C.~M.~Naon and E.~M.~Santangelo,
  Phys.\ Rev.\  D {\bf 33} (1986) 1828;
  K.~Fujikawa,
  Phys.\ Rev.\  D {\bf 29} (1984) 285;
    R.~Jackiw,
CITATION = C84/10/31.

\bibitem{Colladay:1996iz}
  D.~Colladay and V.~A.~Kostelecky,
  Phys.\ Rev.\  D {\bf 55} (1997) 6760
  [arXiv:hep-ph/9703464].

\bibitem{Chu:2000bz}
  C.~S.~Chu,
  Nucl.\ Phys.\  B {\bf 580}, 352 (2000)
  [arXiv:hep-th/0003007];
  A.~P.~Polychronakos,
  JHEP {\bf 0011}, 008 (2000)
  [arXiv:hep-th/0010264];
 G.~H.~Chen and Y.~S.~Wu,
  Nucl.\ Phys.\  B {\bf 628}, 473 (2002)
  [arXiv:hep-th/0111109].
\end{thebibliography}
\end{document}